\documentclass[prc,preprint]{revtex4}

\usepackage{graphicx}

\begin{document}

\title{{\bf
Finite temperature calculations for the spin polarized asymmetric
nuclear matter with the LOCV method }}

\author{{\bf M.
Bigdeli$^{1,3}$\footnote{E-mail: m\underline\ \
bigdeli@znu.ac.ir}},  {\bf G.H. Bordbar$^{2,3}$\footnote{E-mail:
ghbordbar@shirazu.ac.ir}} and {\bf A. Poostforush$^{2}$}}
 \affiliation{
 $^1$Department of Physics, Zanjan University, P.O. Box 45195-313,
Zanjan, Iran\footnote{Permanent address}\\
$^2$Department of Physics, Shiraz University, Shiraz 71454,
Iran\footnote{Permanent address}\\
$^3$Research Institute for Astronomy and Astrophysics of Maragha,\\
P.O. Box 55134-441, Maragha, Iran}


\begin{abstract}
The lowest order constrained variational (LOCV) technique has been
used to investigate some of the thermodynamic properties of spin
polarized hot asymmetric nuclear matter, such as the free energy,
symmetry energy, susceptibility and equation of state.
We have shown that the symmetry energy of the nuclear matter is
substantially sensitive to the value of spin polarization.
 Our calculations show that the equation of state of the polarized hot asymmetric
nuclear matter is stiffer for the higher values of the
polarization as well as the isospin asymmetry parameter.
Our results for the free energy and susceptibility show that the
spontaneous ferromagnetic phase transition cannot occur for hot
asymmetric matter.
\end{abstract}
\maketitle

\section{INTRODUCTION}

The possible occurrence of the spontaneous phase transition to a
ferromagnetic state in nucleonic matter is very important for
studies relevant to astrophysical problems, such as the physical
origin of the magnetic field of the pulsars
\cite{shap,paci,gold,navarro}. The study of the thermodynamic and
magnetic properties of  spin polarized hot asymmetric nuclear
matter, such as free energy, symmetry energy, magnetic
susceptibility plays a crucial role in understanding  the
ferromagnetic phase transition, the equation of state and the
structure of systems as diverse as the neutron rich nuclei and
protoneutron stars.

 A protoneutron star (newborn neutron star) is
born within a short time just after the supernovae collapse. In
this stage, the interior temperature of the neutron star matter is
in the order of 20-50 MeV \cite{burro}. Determination of the
dependence of the symmetry energy on the  spin polarization and
the behavior of the magnetic susceptibility versus the density are
of special interests in  description of the occurrence of the
ferromagnetic phase transition in asymmetric nuclear matter. In
addition these quantities are useful to estimate the mean free
path of the neutrino in the dense nucleonic matter which is a
relevant information for  understanding  the mechanism of the
supernova explosion and the cooling process of the neutron stars
\cite{iwamoto}.

There may exist several possibilities for the generation of the
magnetic field in a neutron star. Among them are the conservation
of the magnetic flux of the original star, a kind of dynamo
mechanism, phase transition to a ferromagnetic state, or a
combination of above mechanisms. The possibility of the existence
of a phase transition to a ferromagnetic state in the neutron
matter and nuclear matter has been studied by several authors
[7-32], without any general agreement.
In most calculations, the neutron star matter is approximated by
pure neutron matter, at zero temperature. Some calculations such
as those based on the hard sphere gas model \cite{brown,rice},
Skyrmelike interactions \cite{vidau}, variational calculation
using the Reid soft-core potential \cite{pandh} and relativistic
Dirac-Hartree-Fock approximation with an effective nucleon-meson
Lagrangian \cite{marcos}, show that the neutron matter becomes
ferromagnetic at some densities. There are other calculations such
as  Monte Carlo \cite{fanto} and Brueckner-Hartree-Fock
calculations \cite{vida,vidab,zuo,zls} using modern two-body and
three-body realistic interactions, which show no indication of
ferromagnetic transition at any density for the neutron matter and
asymmetric nuclear matter.
Properties of the polarized neutron matter at finite temperature
have been  studied by several authors \cite{apv,dapv,bprrv}.
Bombaci et al. \cite{bprrv} have studied these properties  within
the framework of the Brueckner-Hartree-Fock formalism using the
$AV_{18}$ nucleon-nucleon interaction. Their results show no
indication of a ferromagnetic transition at any density and
temperature. Lopez-Val et al. \cite{dapv} have used the D1 and D1P
parameterization of the Gogny interaction and the results of their
calculation show two different behaviors. Whereas the D1P force
exhibits a ferromagnetic transition at a density of $\rho\sim1.31
fm^{-3}$  whose onset increases with temperature, no sign of such
a transition is found for D1 at any densities and temperatures.
Rios et al. \cite{apv} have used Skyrme-like interactions and
their results indicate the occurrence of a ferromagnetic phase of
the neutron matter.
The influence of the finite temperature on the antiferromagnetic
(AFM) spin ordering in the symmetric nuclear matter with the
effective Gogny interaction, within the framework of a Fermi
liquid formalism, has been studied by Isayev \cite{isay1,isay2}.
Here in our article, we use the lowest order constrained
variational (LOCV) formalism to investigate the possibility of the
transition to a ferromagnetic phase for the polarized hot
asymmetric nuclear matter.

The LOCV method has been developed to study the bulk properties of
the quantal fluids \cite{OBI1,OBI2,OBI3}. This technique has been
used for studying the ground state properties of the finite nuclei
and treatment of isobars \cite{BHIM,MI1,MI2}.
 Modarres has extended the LOCV method to the finite temperature
 calculations and has applied it to the neutron matter,
 nuclear matter and asymmetric nuclear matter in order to calculate the
 different thermodynamic properties of these systems \cite{Mod93,Mod95,Mod97,MM98}.
A few years ago, we calculated the properties of
 nuclear matter at zero and finite temperatures using
the LOCV method with the new nucleon-nucleon potentials
\cite{BM97,BM98,MB98}.
Recently we have computed the properties of the spin polarized
neutron matter \cite{bordbig}, the spin polarized symmetric
\cite{bordbig2} and asymmetric nuclear matters and neutron star
matter \cite{bordbig3} at zero temperature. In these works the
microscopic calculations employing the LOCV method with the
realistic nucleon-nucleon potentials have been used. We have
concluded that the spontaneous phase transition to the
ferromagnetic state does not occur.
We have also calculated the thermodynamic properties of the spin
polarized neutron matter \cite{bordbig4} and symmetric nuclear
matter \cite{bordbig5} such as the free energy, magnetic
susceptibility, entropy and pressure using the LOCV method at
finite temperature.
 Our calculations do not show any transition to a ferromagnetic
 phase for hot neutron matter and hot symmetric nuclear matter.

In the present work, we want to  calculate the properties of spin
polarized asymmetric nuclear matter with the LOCV technique at
finite temperature employing the $AV_{18}$ potential
\cite{wiring}.


\section{LOCV calculation of the spin polarized hot asymmetric nuclear matter }

Spin polarized asymmetric nuclear matter is an infinite system
composed of spin-up and spin-down neutrons with densities
$\rho_{n}^{(+)}$ and $\rho_{n}^{(-)}$, respectively, and spin-up
and spin-down protons with densities $\rho_{p}^{(+)}$ and
$\rho_{p}^{(-)}$, respectively. The total densities for neutrons
($\rho_n$) and protons ($\rho_{p}$) are given by:
\begin{eqnarray}
     \rho_{p}=\rho_{p}^{(+)}+\rho_{p}^{(-)},\ \ \ \
     \rho_{n}=\rho_{n}^{(+)}+\rho_{n}^{(-)},
 \end{eqnarray}
and the total density of the system is
\begin{eqnarray}
\rho&=&\rho_{p}+\rho_{n}.
 \end{eqnarray}
 Labels (+) and (-) are used
for spin-up and spin-down nucleons, respectively. One can use the
following parameter to identify a given spin polarized state of
the asymmetric nuclear matter,
\begin{eqnarray}
      \delta_{p}=\frac{\rho_{p}^{+}-\rho_{p}^{-}}{\rho_{p}} , \ \ \
      \delta_{n}=\frac{\rho_{n}^{+}-\rho_{n}^{-}}{\rho_{n}}
 \end{eqnarray}
 $\delta_p$ and $\delta_n$ are proton and neutron spin asymmetry parameters,
 respectively.
 These parameters  can have  values in the range of $0.0$
(unpolarized) to $1.0$ (fully polarized). The asymmetry parameter
which describes the isospin asymmetry of the system is defined as,
\begin{eqnarray}
      \beta=\frac{\rho_{n}-\rho_{p}}{\rho}.
 \end{eqnarray}
Pure neutron matter is  totally asymmetric nuclear matter with
$\beta=1$ and the symmetric nuclear matter has $\beta=0$.

To obtain the macroscopic properties of this system, we should
calculate the total free energy per nucleon, $F=E - {\cal T} S$,
where ${\cal T}$ is the temperature, and $E$ and $S$ are the total
energy and entropy per nucleon, respectively. In the case of spin
polarized asymmetric nuclear matter, the free energy per particle
can be calculated by a parabolic approximation resulted from the
charge independence and time-reversal invariance of the
nucleon-nucleon interaction as follows \cite{vidab},
\begin{eqnarray}\label{ase2}
F(\rho,{\cal T},\beta,\delta_n,\delta_p)&=& F_{snucm}(\rho,{\cal
T},\beta=0,\delta_n=0,\delta_p=0)+{F}_{1}(\rho,{\cal
T})(\frac{1+\beta}{2}\delta_n+\frac{1-\beta}{2}\delta_p)^{2}
\nonumber\\
&&+{F}_{2}(\rho,{\cal T})\beta^{2}+{F}_{3}(\rho,{\cal
T})(\frac{1+\beta}{2}\delta_n-\frac{1-\beta}{2}\delta_p)^{2},
 \end{eqnarray}
where  the coefficients ${F}_{1}(\rho,{\cal T})$,
${F}_{2}(\rho,{\cal T})$ and ${F}_{3}(\rho,{\cal T})$ have been
determined in the following way,
\begin{eqnarray}\label{ase2}
{F}_{1}(\rho,{\cal T})&=& {F}(\rho,{\cal
T},\beta=0,\delta_n=1,\delta_p=1)-{F}_{snucm}(\rho,{\cal
T},\beta=0,\delta_n=0,\delta_p=0),\nonumber\\
{F}_{2}(\rho,{\cal T})&=& {F}(\rho,{\cal
T},\beta=1,\delta_n=0,\delta_p=0)-{F}_{snucm}(\rho,{\cal
T},\beta=0,\delta_n=0,\delta_p=0),\nonumber\\
 {F}_{3}(\rho,{\cal T})&=& {F}(\rho,{\cal
T},\beta=1,\delta_n=1,\delta_p=1)-{F}_{snucm}(\rho,{\cal
T},\beta=0,\delta_n=0,\delta_p=0)\nonumber\\ &&
-{F}_{1}(\rho,{\cal T})-{F}_{2}(\rho,{\cal T}).
 \label{asym}
 \end{eqnarray}

We calculate the total energy per nucleon ($E$) using LOCV method
as follows \cite{bordbig4, bordbig5}.
We adopt a trial many-body wave function of the form
\begin{eqnarray}
     \psi=\cal{F}\phi,
 \end{eqnarray}
where $\phi$ is the uncorrelated ground state wave function of $A$
independent nucleons (simply the Slater determinant of the plane
waves)  and ${\cal F}={\cal F}(1\cdots A)$ is an appropriate
A-body correlation operator which can be replaced by a Jastrow
form i.e.,
\begin{eqnarray}
    {\cal F}={\cal S}\prod _{i>j}f(ij),
 \end{eqnarray}
in which ${\cal S}$ is a symmetrizing operator.
Now, we consider the cluster expansion of the energy functional up
to the two-body term \cite{clark},
 \begin{eqnarray}\label{tener}
           E([f])=\frac{1}{A}\frac{\langle\psi|H\psi\rangle}
           {\langle\psi|\psi\rangle}=E _{1}+E _{2}\cdot
 \end{eqnarray}
For the polarized hot asymmetric nuclear matter, the one-body term
$E _{1}$ is
\begin{eqnarray}\label{ener1}
 E _{1}= \sum_{j=p,n}\ \sum_{i=+,-}E_{1j}^{(i)},
\end{eqnarray}

where $E_{1j}^{(i)}$ is the one-body energy of nucleon $j$ with
spin projection $i$,
 \begin{eqnarray}
E_{1j}^{(i)}=\sum _{k} \frac{\hbar^{2}{k^2}}{2m}n_j^{(i)}(k,{\cal
T},\rho_j^{(i)}).
\end{eqnarray}
 $n_j^{(i)}(k,{\cal T},\rho_j^{(i)})$ is the Fermi-Dirac
distribution function,
\begin{eqnarray}
n_j^{(i)}(k,{\cal
T},\rho_j^{(i)})=\left(e^{\beta\,[\epsilon_j^{(i)} (k,{\cal
T},\rho_j^{(i)})-\mu_j^{(i)}({\cal
T},\rho_j^{(i)})\,]\,}+1\right)^{-1}\cdot
\end{eqnarray}
In the above equation $\beta=\frac{1}{k_B{\cal T}}$ ,
$\mu_j^{(i)}$ is the chemical potential of nucleon $j$ with spin
projection $i$ which is determined at any values of the
temperature ($\cal T$), number density ($\rho_j^{(i)}$) and
polarization ($\delta_j$) by applying the following constraint,
 \begin{eqnarray}\label{chpt}
 \sum _{k}
 n_j^{(i)}(k,{\cal T},\rho_j^{(i)})=A_j^{(i)},
 \end{eqnarray}
and
 $\epsilon_j^{(i)}$ is the single particle energy of nucleon $j$ with
spin projection $i$. In our formalism, the single particle energy
of nucleon $j$ with momentum $k$ and spin projection $i$ is
written approximately in terms of the effective mass as follows
\cite{apv,dapv,isay2}
 \begin{eqnarray}
 \epsilon_j^{(i)}(k,{\cal T},\rho_j^{(i)})=
 \frac{\hbar^{2}{k^2}}{2{m^{*}}_j^{(i)}(\rho,{\cal T})}+U_j^{(i)}({\cal
 T},\rho_j^{(i)}).
                \end{eqnarray}
$U_j^{(i)}({\cal
 T},\rho_j^{(i)})$ is the
momentum independent single particle potential. In fact, we use a
quadratic approximation for single particle potential incorporated
in the single particle energy as a momentum independent effective
mass and introduce the effective masses, $m_j^{{*}{(i)}}$, as
variational parameters \cite{bordbig4,fp}. We minimize the free
energy with respect to the variations in the effective masses and
then obtain the chemical potentials and the effective masses of
the spin-up and spin-down nucleons at the minimum point of the
free energy. This minimization is done numerically.

The two-body energy $E_{2}$ is
\begin{eqnarray}
    E_{2}&=&\frac{1}{2A}\sum_{ij} \langle ij\left| \nu(12)\right|
    ij-ji\rangle,
\end{eqnarray}
 where
\begin{eqnarray}
 \nu(12)=-\frac{\hbar^{2}}{2m}[f(12),[\nabla
_{12}^{2},f(12)]]+f(12)V(12)f(12).
\end{eqnarray}
In above equation, $f(12)$ and $V(12)$ are the two-body
correlation and potential.
In our calculations, we use the $AV_{18}$ two-body potential which
has the following form \cite{wiring},
\begin{equation}
V(12)=\sum^{18}_{p=1}V^{(p)}(r_{12})O^{(p)}_{12}, \label{v18}
\end{equation}
where
\begin{eqnarray}
O_{12}^{(p=1-18)}&=&1,\ {\bf\sigma_1}\cdot{\bf\sigma_2},\
{\bf\tau_1}\cdot{\bf\tau_2},\ ({\bf\sigma_1}\cdot{\bf\sigma_2})\
({\bf\tau_1}\cdot{\bf\tau_2}),\ S_{12},\
S_{12}({\bf\tau_1}\cdot{\bf\tau_2}),
\nonumber\\
&&{\bf L}\cdot{\bf S},\ {\bf L}\cdot {\bf
S}({\bf\tau_1}\cdot{\bf\tau_2}),\ {\bf L}^2,\ {\bf
L}^2({\bf\sigma_1}\cdot{\bf\sigma_2}),\
{\bf L}^2({\bf\tau_1}\cdot{\bf\tau_2}),\nonumber\\
&&{\bf
L}^2({\bf\sigma_1}\cdot{\bf\sigma_2})({\bf\tau_1}\cdot{\bf\tau_2}),\
({\bf L}\cdot {\bf S})^2,\ ({\bf L}\cdot {\bf
S})^2({\bf\tau_1}\cdot{\bf\tau_2}),\nonumber\\
&&{\bf T_{12}},\ ({\bf\sigma_1}\cdot{\bf\sigma_2}){\bf T_{12}},\
S_{12}{\bf T_{12}},\ (\bf\tau_{z1}+\bf\tau_{z2}). \label{operat}
\end{eqnarray}
In above equation, $S_{12}=[3({\bf\sigma_1}\cdot\hat
r)({\bf\sigma_2}\cdot\hat r)-{\bf\sigma_1}\cdot{\bf\sigma_2}]$ is
the tensor operator and ${\bf T_{12}}=[3({\bf\tau_1}\cdot\hat
r)({\bf\tau_2}\cdot\hat r) -{\bf\tau_1}\cdot{\bf\tau_2}]$ is the
isotensor operator \cite{wiring}.
In the LOCV formalism, the two-body correlation $f(12)$ is
considered to have the following form \cite{OBI3},
\begin{eqnarray}
f(12)&=&\sum^3_{k=1}f^{(k)}(r_{12})P^{(k)}_{12},
\end{eqnarray}
where
\begin{eqnarray}
P_{12}^{(k=1-3)}&=&\left (\frac{1}{4} -
\frac{1}{4}O^{(2)}_{12}\right ), \ \left (\frac{1}{2} +
\frac{1}{6}O^{(2)}_{12} + \frac{1}{6}O^{(5)}_{12}\right
),\nonumber\\&&\left (\frac{1}{4} + \frac{1}{12}O^{(2)}_{12} -
\frac{1}{6}O^{(5)}_{12}\right ).
\end{eqnarray}
The operators $O^{(2)}_{12}$ and $O^{(5)}_{12}$ are given in Eq.
(\ref{operat}).

In our formalism, we  minimize the two-body energy $E_2$ with
respect to the variations in the correlation functions
${f_{\alpha}}^{(k)}$ subject to the normalization constraint
\cite{OBI3,BM98},
\begin{eqnarray}\label{norm}
        \frac{1}{A}\sum_{ij}\langle ij\left| h_{S_{z}}^{2}
        -f^{2}(12)\right| ij\rangle _{a}=0\cdot
 \end{eqnarray}
In the case of polarized nuclear matter, the Pauli function
$h_{S_{z}}(r)$ is as follows

\begin{eqnarray}
h_{S_{z}}(r)=
\left\{%
\begin{array}{ll}
\left[ 1-\frac{1}{\nu}\left( \frac{\gamma^{(i)}(r)
       }{\rho}\right) ^{2}\right] ^{-1/2} & ;\ \hbox{$S_{z}=\pm 1$} \\
    1 & ;\ \hbox{$S_{z}= 0$}
\end{array}%
\right.
\end{eqnarray}
 where
\begin{eqnarray}
\gamma^{(i)}(r)=\frac{2\nu}{(2\pi)^{2}}\int n^{(i)}(k,{\cal
T},\rho^{(i)})J_{0}(kr)k^2dk .
 \end{eqnarray}
Here $\nu$ is the degeneracy of the system. From the minimization
of the two-body cluster energy, we get a set of coupled and
uncoupled differential equations the same as presented in Ref.
\cite{bordbig5}. We can obtain the correlation functions by
solving these differential equations and then calculate the
two-body energy. Finally, we can compute the energy and the free
energy of the system.

\section{Results and discussion }\label{NLmatchingFFtex}
\begin{figure}

\includegraphics[width=8.1cm]{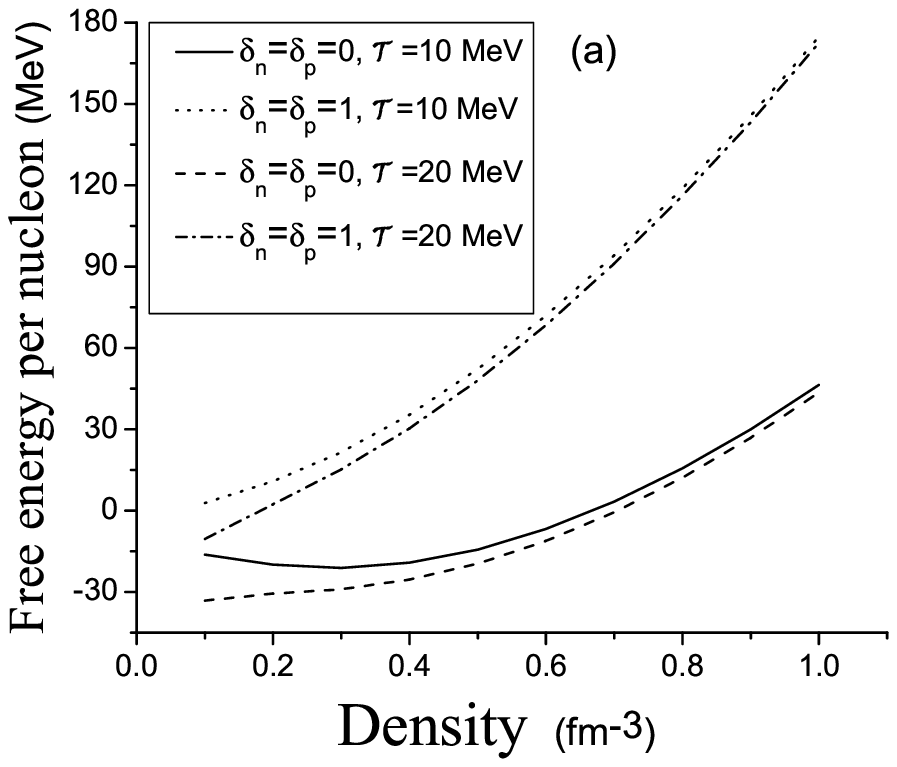}
\includegraphics[width=8.1cm]{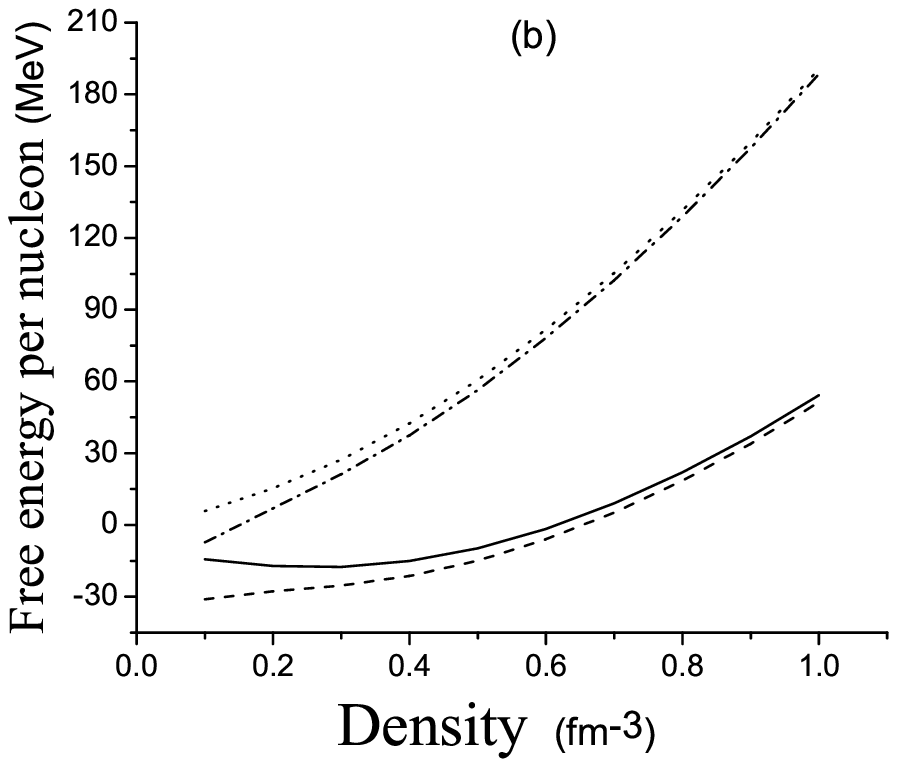}
\includegraphics[width=8.1cm]{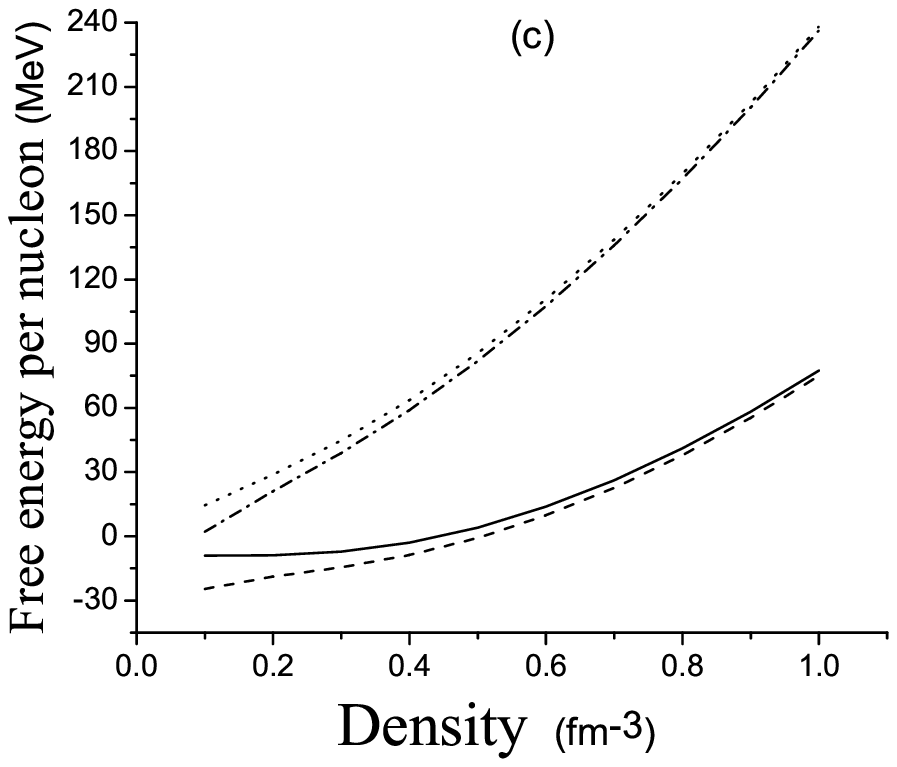}
\includegraphics[width=8.1cm]{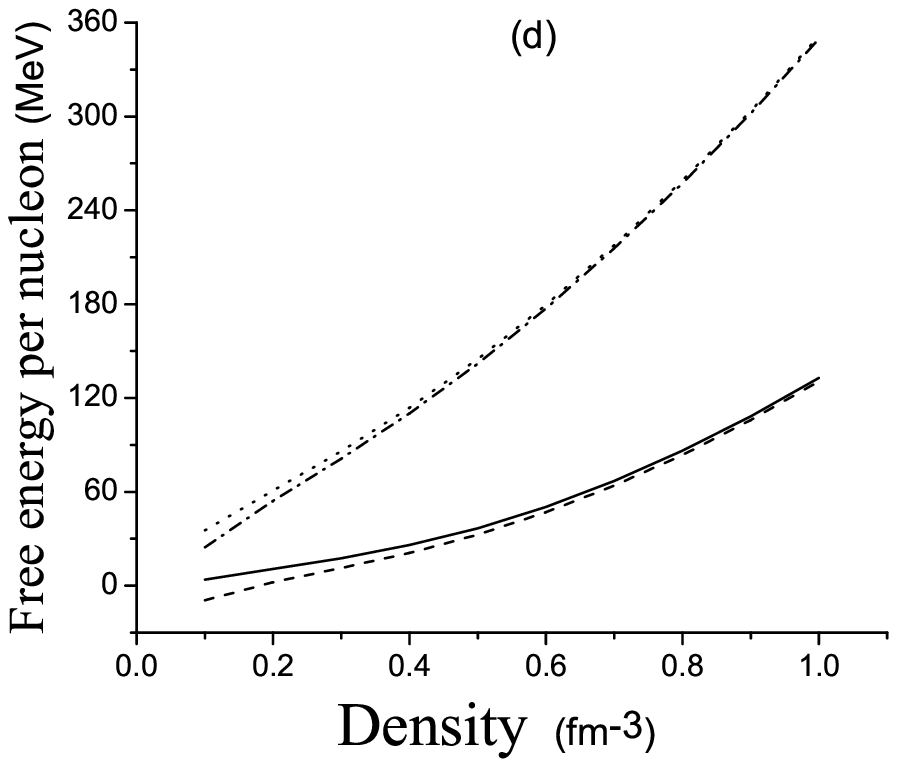}

 \caption{The free energy
per nucleon of spin polarized hot asymmetric nuclear matter as a
function of the total number density ($\rho$) for unpolarized and
fully polarized matter at  ${\cal T}=10$ and $20\ $MeV for
$\beta=0.0$ (a), $0.3$ (b), $0.6$ (c) and $1.0$
(d).}\label{freepol1}
\end{figure}

Fig. \ref{freepol1} shows the free energy per nucleon of the spin
polarized hot asymmetric nuclear matter versus the total number
density ($\rho$) for different values of neutron polarization
($\delta_n$), proton polarization ($\delta_p$) and isospin
asymmetry parameter ($\beta$) at ${\cal T}=10 $ and $20\ MeV$. It
can be seen from this figure that at each temperature for a given
value of isospin asymmetry parameter, the free energy of polarized
hot asymmetric nuclear matter increases by increasing the
polarization.  For all temperatures and isospin asymmetry
parameters, we do not see any crossing between the free energy
curves of different polarizations. The difference between the free
energies of different polarizations increases by increasing the
density. Therefore, we can conclude that there is no spontaneous
transition to the ferromagnetic phase for the hot asymmetric
nuclear matter.
Fig. \ref{freepol1}  also shows that for a given polarization and
temperature, the free energy increases by increasing the isospin
asymmetry parameter.
 From Fig. \ref{freepol1}, we see that the free energy
of spin polarized asymmetric nuclear matter decreases by
increasing the temperature.
We have found that only for the lower values of temperature
(${\cal T}$), neutron polarization ($\delta_n$), proton
polarization ($\delta_p$) and isospin asymmetry parameter
($\beta$), the free energy curve shows a minimum point. However,
at higher values of these quantities, this minimum point
disappears showing no bound state for spin polarized hot
asymmetric nuclear matter.

\begin{figure}

\includegraphics{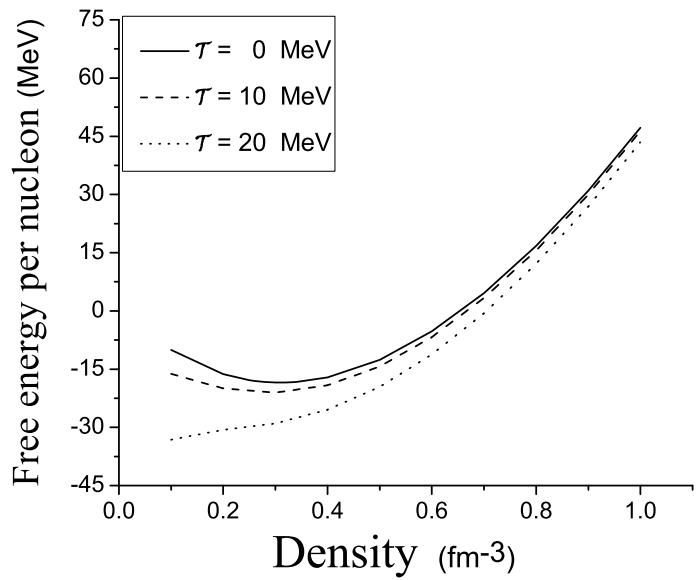}

 \caption{The free energy
per nucleon of the unpolarized hot symmetric nuclear matter
($\beta=0.0$) versus the total number density ($\rho$) at ${\cal
T}=0,\ 10$ and $20\ MeV$.} \label{freeunpol}
\end{figure}

For the unpolarized case of nuclear matter with $\beta=0.0$
(symmetric nuclear matter), we have compared the free energies at
different temperatures in Fig. \ref{freeunpol}. It is seen that
the free energy of unpolarized symmetric nuclear matter decreases
by increasing the temperature, especially at low densities. From
Fig. \ref{freeunpol}, we can see that at zero temperature, the
saturation density (density of minimum point of energy) for the
unpolarized symmetric nuclear matter is about $\rho=0.31\ fm^{-3}$
which is greater than the empirical value ($0.16\ fm^{-3}$). Our
results also show that at ${\cal T}=0\ MeV$,  the energy of
symmetric nuclear matter at our calculated saturation density
(binding energy) is about $-18\ MeV$ which is smaller than the
empirical value ($-16\ MeV$).

\begin{figure}

\includegraphics{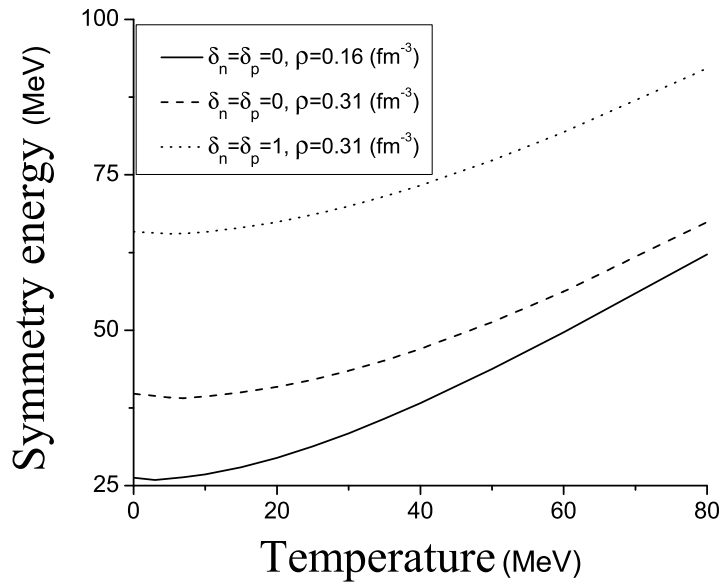}

 \caption{The symmetry energy
 of the spin polarized hot asymmetric nuclear matter versus
temperature for two values of the density at different
polarizations.} \label{symenergy}
\end{figure}

\begin{figure}

\includegraphics{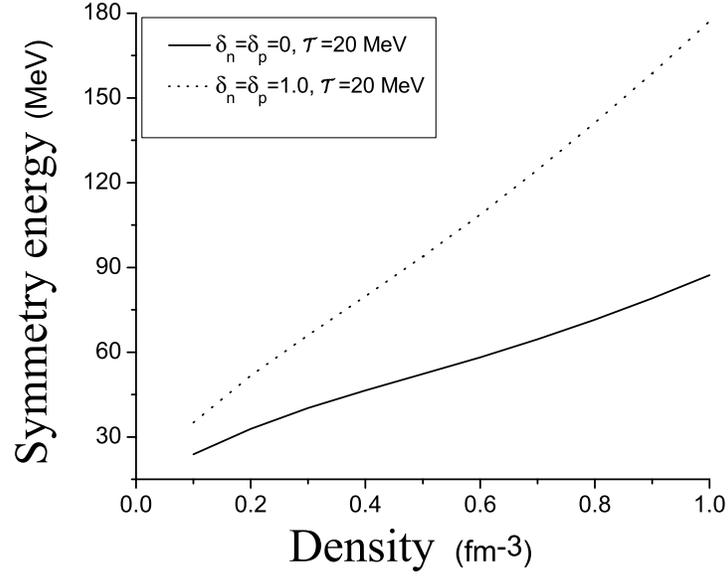}

 \caption{The symmetry energy
 of spin polarized hot asymmetric nuclear matter versus
 density at ${\cal T}= 20\ MeV$ for different polarizations.
} \label{symenergy2}
\end{figure}

For the polarized hot asymmetric nuclear matter, the nuclear
symmetry energy is given by
\begin{eqnarray}\label{ase2}
F_{sym}(\rho,{\cal T},\delta_n,\delta_p)&=& {F}(\rho,{\cal
T},\beta=1,\delta_n,\delta_p)-{F}(\rho,{\cal
T},\beta=0,\delta_n,\delta_p).
 \label{asym}
 \end{eqnarray}
In Fig. \ref{symenergy}, we have plotted the nuclear symmetry
energy (${F}_{sym}$) of  spin polarized hot asymmetric nuclear
matter as a function of the temperature at $\rho=0.31$ and $0.16\
fm^{-3}$ for two cases with $\delta_n=\delta_p=0.0$ and
$\delta_n=\delta_p=1.0$.
It is seen that at low temperatures, for a given polarization, the
difference between the symmetry energies of different densities is
substantially large. However, this difference decreases as the
temperature increases.
Fig. \ref{symenergy} shows that for each density, the symmetry
energy increases by increasing the polarization and temperature.
However, for each density, the difference between the symmetry
energies of different polarization decreases by increasing the
temperature.
From Fig. \ref{symenergy}, for the case of unpolarized nuclear
matter ($\delta_n=\delta_p=0$), at ${\cal T}=0\ MeV$, we have
found that the value of the symmetry energy corresponding to our
calculated saturation density ($\rho=0.31\ fm^{-3}$) is about $39\
MeV$.
For this case of nuclear matter, the empirical value of the
nuclear symmetry energy is $28-32\ MeV$ corresponding to the
empirical saturation density of about $\rho=0.16\ fm^{-3}$
\cite{haus}.
This shows that our calculated symmetry energy as well as our
saturation density are greater than the empirical results.
We have also found that the symmetry energy of the unpolarized
nuclear matter at ${\cal T}=0\ MeV$ for $\rho=0.16\ fm^{-3}$ is
about $26\ MeV$, which is less than its empirical value.
The density dependence of the nuclear symmetry energy of polarized
hot asymmetric nuclear matter has been also shown in Fig.
\ref{symenergy2}. For each value of the polarization, we see that
${F}_{sym}$ is an increasing function of the density. However, the
rate of increasing of the symmetry energy versus the density
increases by increasing the polarization.

\begin{figure}

\includegraphics{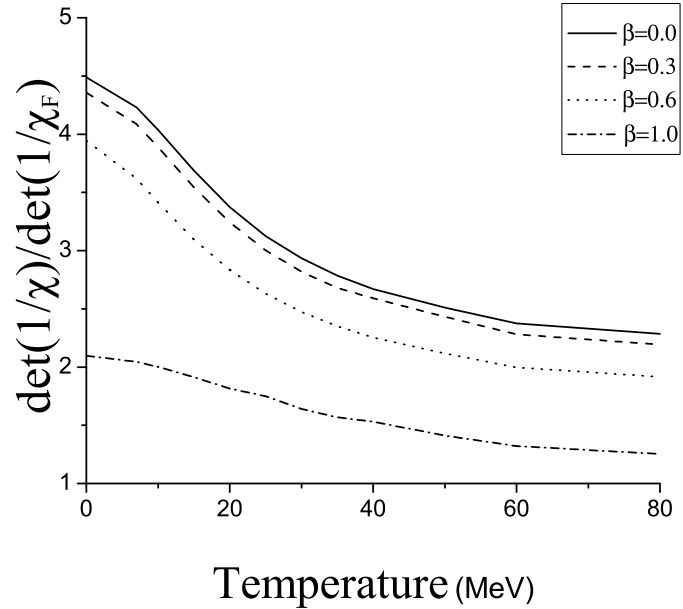}

 \caption{The magnetic susceptibility of hot asymmetric nuclear matter
versus temperature at $\rho=0.16 fm^{-3}$ for different values of
the asymmetry parameter ($\beta$) .} \label{sustem}
\end{figure}

\begin{figure}

\includegraphics{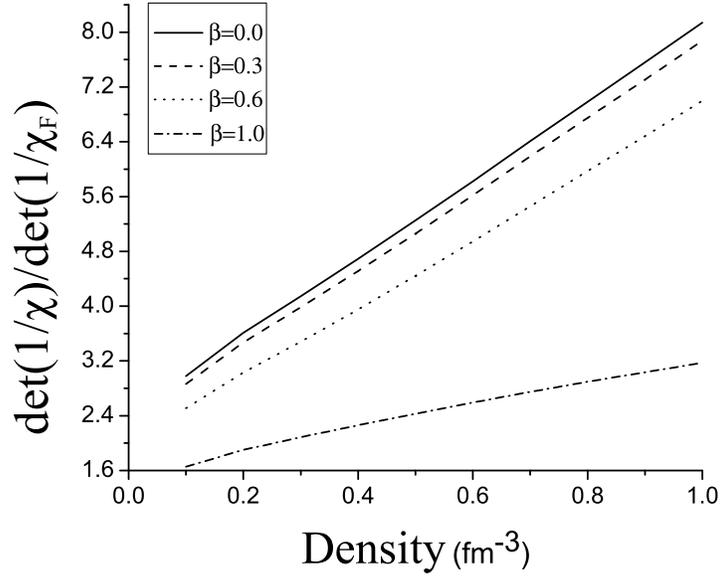}

 \caption{The magnetic
susceptibility of hot asymmetric nuclear matter versus the total
number density ($\rho$) at ${\cal T}=20$ MeV for different values
of the asymmetry parameter ($\beta$).} \label{susden}
\end{figure}

The response of a system to the magnetic field is characterized by
the magnetic susceptibility, $\chi$, which in the case of
asymmetric nuclear matter is defined by a $2\times2$ matrix as
follows,
\begin{eqnarray}\label{susep}
   1/\chi&=&\left(%
\begin{array}{cc}
  1/\chi_{nn} & 1/\chi_{np} \\
  1/\chi_{pn} & 1/\chi_{pp} \\
\end{array}%
\right),
\end{eqnarray}
where the matrix elements $1/\chi_{ij}$ are given by,

\begin{eqnarray}\label{susep}
   1/\chi_{ij} =\frac{\rho}{\mu_{i}\rho_i\mu_j\rho_j}
   \left( \frac{\partial^{2}F}{\partial \delta_{i}\partial \delta_j}\right)
_{\delta_i=\delta_j =0}.
\end{eqnarray}
In the above equation, $\rho$ is the total density, $\mu_i$
($\mu_j$) and $\rho_i$ ($\rho_j$) are the magnetic moment and
density of the particle $i$ ($j$), respectively, and $\delta_{i}$
and $ \delta_j$ are the spin asymmetry parameters of particles $i$
and $j$, respectively.
The onset of the spin instability of the system appears when the
sign of the determinant of this matrix becomes negative.
We calculate the magnetic susceptibility of the polarized
asymmetric nuclear matter in terms of the ratio
$\frac{\det(1/\chi)}{\det(1/\chi_F)}$, where $\chi_{F}$ is the
magnetic susceptibility for an ideal Fermi gas containing
noninteracting protons and neutrons.
For the polarized hot asymmetric nuclear matter, the ratio
$\frac{\det(1/\chi)}{\det(1/\chi_F)}$ versus the temperature, for
$\rho=0.16 fm^{-3}$ at different values of the isospin asymmetry
parameter ($\beta$) has been presented in Fig. \ref{sustem}.
This figure shows that for all values of the isospin asymmetry
parameter and temperature, $\frac{\det(1/\chi)}{\det(1/\chi_F)}$
is positive. This behavior indicates that there is no magnetic
instability for the hot asymmetric nuclear matter.
From Fig. \ref{sustem}, we see that for all  isospin asymmetry
parameters, the ratio $\frac{\det(1/\chi)}{\det(1/\chi_F)}$ which
is greater than unity, monotonically decreases by increasing the
temperature.
This shows that as the temperature increases, the strong
correlation in nucleonic matter which arises by nucleon-nucleon
interaction, becomes less important.
Fig. \ref{sustem} also shows that for all given temperatures, the
ratio $\frac{\det(1/\chi)}{\det(1/\chi_F)}$ increases by
decreasing the asymmetry parameter ($\beta$). This indicates that
the proton fraction substantially affects the magnetic
susceptibility of asymmetric nuclear matter. This effect has been
already pointed out by Kutschera et al. \cite{kw} and Bernardos et
al. \cite{bern}.
In Fig. \ref{susden}, we have plotted the ratio
$\frac{\det(1/\chi)}{\det(1/\chi_F)}$ as a function of the density
at ${\cal T}=20\ MeV$ for a wide range of the isospin asymmetry
parameters. We see that for all densities, this ratio  is always
positive, indicating that the spontaneous ferromagnetic phase
transition does not occur for hot asymmetric nuclear matter at any
densities.
Fig. \ref{susden} also shows that this ratio is always greater
than unity and always increases by increasing the density.
Therefore, we can conclude that the strong nucleon-nucleon
correlation becomes more important as the density of system
increases.

\begin{figure}

\includegraphics{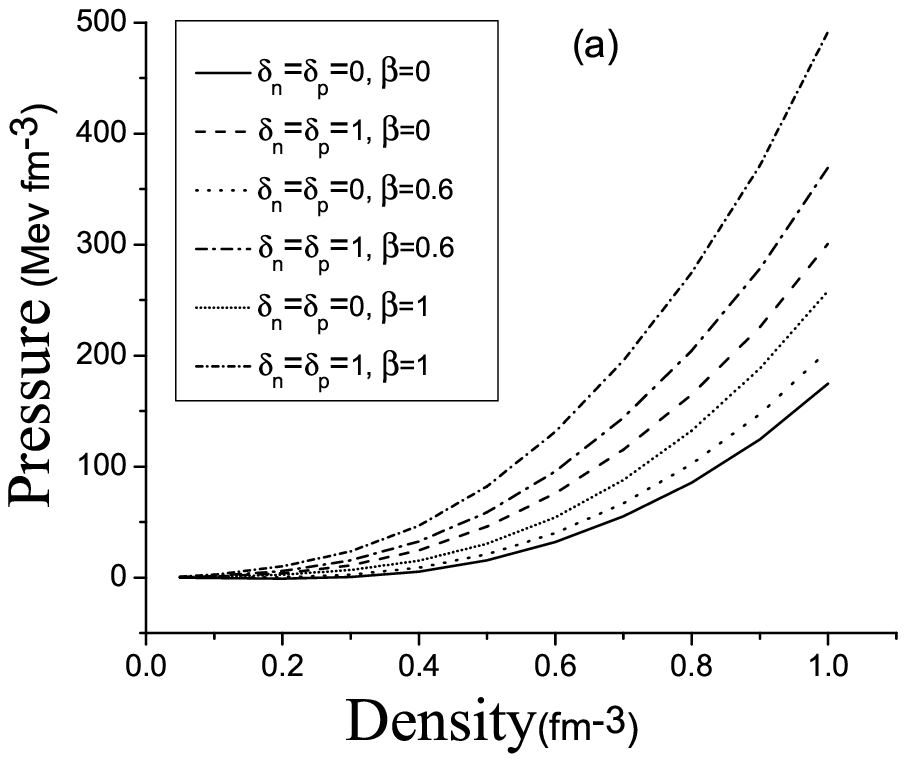}
\includegraphics{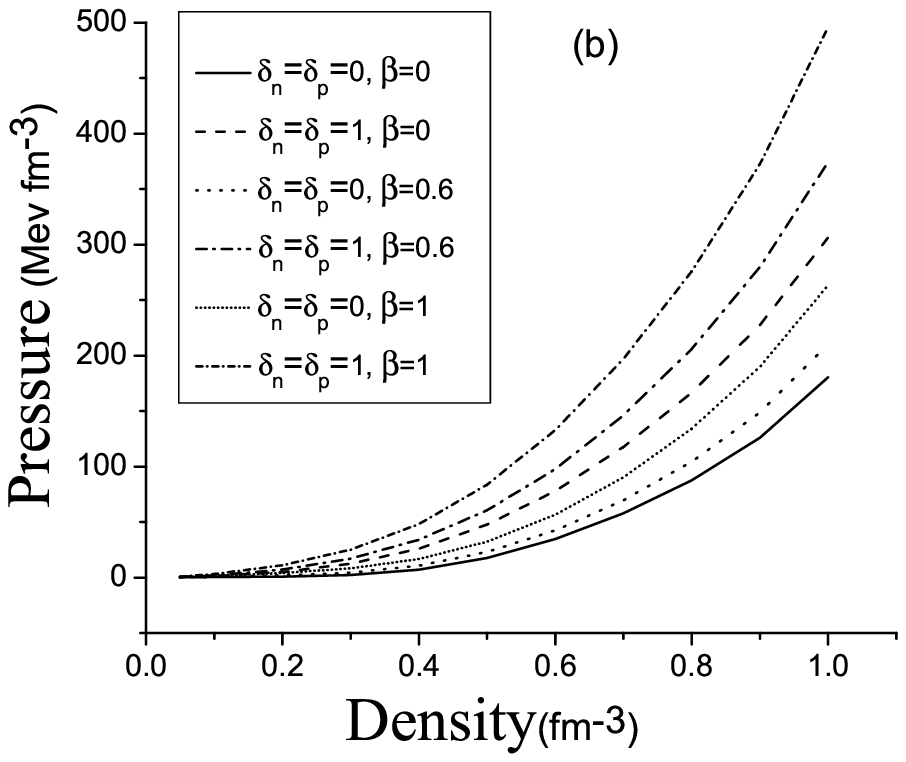}

\caption{The equation of state of spin polarized hot asymmetric
nuclear matter for different values of the  polarization and
asymmetry parameter at ${\cal T}=10$ MeV (a) and ${\cal T}=20$ MeV
(b).} \label{pres1}
\end{figure}

The equation of state of the polarized hot asymmetric nuclear
matter for different values of the polarization and isospin
asymmetry parameter at ${\cal T}=10$ and $20\ MeV$ has been
presented in Fig. \ref{pres1}.
 We see that for all  temperatures and isospin asymmetry parameters,
 the equation of state becomes stiffer as the polarization
increases. We  also see that for a given  polarization at each
temperature, the pressure of spin polarized hot asymmetric nuclear
matter increases by increasing the isospin asymmetry parameter.
By comparing the two panels of Fig. \ref{pres1}, we can see that
increasing the temperature leads to the stiffer equation of state
for spin polarized asymmetric nuclear matter.
\section{Summary and Conclusions}
We have used the lowest order constrained variational (LOCV)
method to calculate the free energy of the polarized hot
asymmetric nuclear matter at different temperatures, for various
values of the isospin asymmetry parameter and spin polarization
employing the $AV_{18}$ two-nucleon potential. Our results show
that the free energy of this system increases by increasing both
the isospin asymmetry parameter and the spin polarization while it
decreases by increasing the temperature. We have also calculated
the symmetry energy of this system to show that this term depends
on the spin  polarization, temperature and density.
 The magnetic susceptibility of  spin
polarized hot asymmetric nuclear matter has been also calculated
for a wide range of  densities and isospin asymmetry parameters at
different temperatures. Our calculations do not show any
spontaneous phase transition to the ferromagnetic state.
We have also computed the pressure of spin polarized hot
asymmetric nuclear matter for different values of the
polarization, isospin asymmetry parameter and temperature. Our
results show that for all temperatures and densities, the equation
of state of this system is an increasing function of spin
polarization and the isospin asymmetry parameter.
Finally, an agreement is seen between our results and those of
other many-body calculations.


\acknowledgements { This work has been supported by Research
Institute for Astronomy and Astrophysics of Maragha. We wish to
thank Shiraz University and Zanjan University Research Councils.}



\newpage



\newpage


\newpage

\newpage


\newpage


\newpage


\newpage




\begin{thebibliography}{99}
\bibitem{shap} S. Shapiro and S. Teukolsky, Blak Holes, White Dwarfs
and Neutron Stars, (Wiley-New York, 1983).
\bibitem{paci} F. Pacini, {Nature} (London) {\bf 216} (1967) 567.
\bibitem{gold} T. Gold, {Nature} (London) {\bf 218} (1968) 731.
\bibitem{navarro}J. Navarro, E. S. Hern´andez and D. Vautherin,
{Phys. Rev}. {\bf C 60} (1999) 045801.
\bibitem{burro} A. Burrows and J. M. Lattimer, {Astrophys. J.}
{\bf 307} (1968) 178.
\bibitem{iwamoto} N. Iwamoto and C. J. Pethick {Phys. Rev}. {\bf D 25} (1982) 313.
\bibitem{brown} D. H. Brownell and J. Callaway, {Nuovo Cimento}
{\bf B 60} (1969) 169.
\bibitem{rice} M. J. Rice, {Phys. Lett}. {\bf A 29} (1969) 637.
\bibitem{clark} J. W. Clark and N. C. Chao, {Lettere Nuovo Cimento}
{\bf 2} (1969) 185.
\bibitem{clark2} J. W. Clark, {Phys. Rev. Lett}. {\bf 23} (1969) 1463.
\bibitem{silv} S. D. Silverstein, {Phys. Rev}. Lett. {\bf 23} (1969) 139.
\bibitem{ostga} E. {\O}stgaard, {Nucl. Phys}. {\bf A 154} (1970) 202.
\bibitem{pear} J. M. Pearson and G. Saunier, {Phys. Rev. Lett}. {\bf 24}
(1970) 325.
\bibitem{pandh} V. R. Pandharipande, V. K. Garde and J. K. Srivastava,
{Phys. Lett}. {\bf B 38} (1972) 485.
\bibitem{backm}  S. O. Backman and C. G. Kallman, {Phys. Lett}.
{\bf B 43} (1973) 263.
\bibitem{haens} P. Haensel, {Phys. Rev}. {\bf C 11} (1975) 1822.
\bibitem{jack}  A. D. Jackson, E. Krotscheck, D. E. Meltzer and
R. A. Smith, {Nucl. Phys}. {\bf A 386} (1982) 125.
\bibitem{kuts} M. Kutschera and W. W´ojcik, {Phys. Lett}.
{\bf B 223} (1989) 11.
\bibitem{marcos} S. Marcos, R. Niembro, M. L. Quelle and J. Navarro,
{Phys. Lett}. {\bf B 271} (1991) 277.
\bibitem{bern} P. Bernardos, S. Marcos, R. Niembro and M. L. Quelle,
{Phys. Lett}. {\bf B 356} (1995) 175.
\bibitem{vidau} A. Vidaurre, J. Navarro and J. Bernabeu, {Astron. Astrophys}.
{\bf 35} (1984) 361.
\bibitem{kutsb} M. Kutschera and W. W´ojcik, {Phys. Lett}. {\bf B 325} (1994) 271.
\bibitem{fanto}S. Fantoni, A. Sarsa and K. E. Schmidt, {Phys. Rev. Lett}.
{\bf 87} (2001) 181101.
\bibitem{vida}I. Vida˜na, A. Polls and A. Ramos, {Phys. Rev}. {\bf C 65}
(2002) 035804.
\bibitem{vidab} I. Vida˜na and I. Bombaci, {Phys. Rev}. {\bf C 66} (2002) 045801.
\bibitem{zuo}W. Zuo, U. Lombardo and C.W. Shen, in Quark-Gluon Plasma and
Heavy Ion Collisions, Ed. W.M. Alberico, M. Nardi and M.P.
Lombardo, World Scientific, p. 192 (2002).
\bibitem{zls} W. Zuo, U.
Lombardo and C. W. Shen,
nucl-th/0204056.\\
W. Zuo, C. W. Shen and U. Lombardo, {Phys. Rev.} {\bf C 67} (2003)
037301.
\bibitem{isay}A. A. Isayev and J. Yang, {Phys. Rev}. {\bf C 69} (2004) 025801.
\bibitem{apr} A. Akmal, V. R. Pandharipande and D. G. Ravenhall,
Phys. Rev. {\bf C 58} (1998) 1804.
\bibitem{apv} A.Rios, A. Polls and I. Vidana, {Phys. Rev.} {\bf C
71}(2005) 055802.
\bibitem{dapv} D. Lopez-Val, A.Rios, A. Polls and I. Vidana, {Phys. Rev.} {\bf C
74} (2006) 068801.
\bibitem{bprrv} I. Bombaci, A. Polls, A. Ramos, A.Rios and I. Vidana, {Phys.
Lett.} {\bf B 632} (2006) 638.

\bibitem{isay1}A. A. Isayev, {Phys. Rev.} {\bf C 72} (2005) 014313.
\bibitem{isay2}A. A. Isayev, {Phys. Rev.} {\bf C 76} (2007) 047305.
{\bf B 632} (2006) 638.



\bibitem{OBI1} J. C. Owen, R. F. Bishop and J. M. Irvine, {Ann. Phys.,
N.Y.}, {\bf 102}, 170 (1976).

\bibitem{OBI2} J. C. Owen, R. F. Bishop and J. M. Irvine,
 {Nucl. Phys.} {\bf A 274}, 108 (1976).

\bibitem{OBI3} J. C. Owen, R. F. Bishop and J. M. Irvine,
{Nucl. Phys.} {\bf A 277}, 45 (1977).

\bibitem{BHIM} R. F. Bishop, C. Howes, J. M. Irvine and M. Modarres,
{J. Phys. G: Nucl. Phys.} {\bf 4}, 1709 (1978).
\bibitem{MI1}  M. Modarres and J. M. Irvine , {J. Phys. G: Nucl. Phys.} {\bf 5}, 7
(1979).
\bibitem{MI2} M. Modarres and J. M. Irvine , {J. Phys. G: Nucl.
Phys.} {\bf 5}, 511 (1979).

\bibitem{Mod93} M. Modarres, {J. Phys. G: Nucl. Phys.} {\bf 19}, 1349 (1993).
\bibitem{Mod95} M. Modarres, {J. Phys. G: Nucl. Phys.} {\bf 21}, 351 (1995).
\bibitem{Mod97} M. Modarres, {J. Phys. G: Nucl. Phys.} {\bf 23}, 923 (1997).
\bibitem{MM98} H. R. Moshfegh and M. Modarres, {J. Phys. G: Nucl. Phys.} {\bf 24}, 821 (1998).


\bibitem{BM97} G. H. Bordbar and M. Modarres, {J. Phys. G: Nucl. Phys.}
{\bf 23}, 1631 (1997).
\bibitem{BM98} G. H. Bordbar and M. Modarres, {Phys. Rev.} {\bf C 57}, 714 (1998).
\bibitem{MB98} M. Modarres and G. H. Bordbar, {Phys. Rev.} {\bf C 58}, 2781 (1998).

\bibitem{bordbig}G. H. Bordbar and M. Bigdeli, {Phys. Rev.}
{\bf C 75} (2007) 045804.
\bibitem{bordbig2}G. H. Bordbar and M. Bigdeli, {Phys.
Rev.} {\bf C 76} (2007) 035803.
\bibitem{bordbig3}G. H. Bordbar and M. Bigdeli, {Phys.
Rev.} {\bf C 77} (2008) 015805.
\bibitem{bordbig4}G. H. Bordbar and M. Bigdeli, {Phys.
Rev.} {\bf C 78} (2008) 054315.
\bibitem{bordbig5} M. Bigdeli, G. H. Bordbar and Z. Rezaei {Phys.
Rev.} {\bf C 80} (2009) 034310.

\bibitem{wiring}R. B. Wiringa, V. Stoks and R. Schiavilla,
{Phys. Rev.} {\bf C 75} (1995) 38.


\bibitem{fp} B. Friedman and V.R. Pandharipande, {Nucl. Phys.} {\bf A
361} (1981) 502.
\bibitem{bomb1} I. Bombaci and U. Lombardo, Phys. Rev. {\bf C 44} (1991) 1892.
\bibitem{li}B. A. Li, {Phys.
Rev.} {\bf C 69} (2004) 064602.

\bibitem{haus} P. E. Haustein, At. Data
Nucl. Data Tables {\bf 39} (1988) 185.

\bibitem{kw} M. Kutschera and W. W´ojcik, {Phys. Lett}. {\bf B 223} (1989) 11.

\end{thebibliography}
\end{document}